\documentclass[12pt, draftclsnofoot, onecolumn]{IEEEtran} 

\usepackage{cite}

\ifCLASSINFOpdf
 \else
    \usepackage[dvips]{graphicx}
     \DeclareGraphicsExtensions{.eps}
\fi

\usepackage{epsfig,amssymb}
\usepackage{epstopdf}

\usepackage[cmex10]{amsmath}
\usepackage{array}
\usepackage[tight,footnotesize]{subfigure}
\usepackage{multirow}
\usepackage{dcolumn}
\usepackage{amssymb}
\usepackage{color}
\usepackage{color}
\def\one{\mbox{1\hspace{-4.25pt}\fontsize{12}{14.4}\selectfont\textrm{1}}}
\begin{document}

\title{Outage Probability of Millimeter Wave  Cellular Uplink with Truncated Power Control}

\author{
Oluwakayode Onireti,~\IEEEmembership{Member,~IEEE,}
        Lei Zhang,~\IEEEmembership{Member,~IEEE,} Ali Imran,~\IEEEmembership{Member,~IEEE,} and
       Muhammad~Ali~  Imran,~\IEEEmembership{Senior Member,~IEEE}

  }

{}


\maketitle

\begin{abstract}\vspace{-3mm}
In this paper, using the stochastic geometry, we
develop a tractable uplink modeling framework for the outage
probability of both the single and multi-tier millimeter wave (mmWave) cellular
networks.
Each tier's mmWave base stations (BSs) are randomly located and they have particular spatial density,  antenna gain, receiver sensitivity, blockage parameter and pathloss exponents. 
Our model takes account
of the maximum power limitation and the per-user  power control. More specifically, each user, which
could be in line-of-sight (LOS) or non-LOS to its serving mmWave
BS, controls its transmit power such that the received
signal power at its serving BS is equal to a predefined threshold. Hence, a truncated channel inversion power control scheme
is implemented for the uplink of mmWave cellular
networks. We derive closed-form expressions for the signal-to-interference-and-noise-ratio (SINR) outage probability for the uplink of both the single and multi-tier
mmWave cellular networks. Furthermore, we analyze the case with a dense network by utilizing the simplified model, where the LOS region is approximated as a fixed LOS disc. The results show that imposing a maximum power constraint    on the user significantly affects the SINR outage probability in the uplink of mmWave cellular networks.
\end{abstract}

\vspace{-2mm}
\begin{IEEEkeywords}
\vspace{-2mm}
 mmWave, power control, stochastic geometry, truncated channel inversion,  uplink communication.
\end{IEEEkeywords}


\vspace{-3mm}
\section{Introduction}
A  fundamental requirement for the 5G-and-beyond mobile networks is the radical increase in data rate. Recent studies have identified massive multiple-input-multiple-output (MIMO), extreme network densification,  and increased bandwidth  as the key technologies toward meeting this requirement \cite{Andrews2014}. 
The  millimeter wave (mmWave) frequencies (ranging from  $30-300~\mathrm{GHz}$) offers a large available bandwidth thus, making them attractive for the 5G mobile networks \cite{Andrews2014,Zhouyue2011,Rappaport2013}.  
Meanwhile, the mmWave band has long been considered ill-suited for the cellular communication due to the excessive pathloss and the poor penetration through materials such as concrete and water. 
Recent capacity studies and survey measurement on mmWave technologies in  \cite{Rappaport2013,Rappaport2013A,Rangan2014,Akdeniz2014} have shown its great promise for 5G urban small cell deployments. The recent advances in low-power CMOS RF circuit and the smaller wavelength associated with the band have further substantiated this promise. 
The later also makes it viable to have  more miniaturized antennas  within the same physical area of the transmitter and receiver \cite{Zhouyue2011,Akdeniz2014}. 
 Further, with a large antenna array, the mmWave network can apply beamforming at the transmit and
receive sides to provide array gain which  compensates for the pathloss\cite{Roh2014}. The directionality gained from beamforming will lead to a reduction in  interference 
\cite{Rangan2014}. Hence,  mmWave spectrum holds great potential for providing the high data rate (Gigabits range) expected in the upcoming 5G cellular networks \cite{Bai2015}.

Modeling and analysis of cellular networks by using stochastic geometry have recently received significant attention due to its high accuracy and tractability. In this approach, the network topology is abstracted to a point process for ease of modeling and analysis. Earlier works in this area were mainly focused on the conventional ultra-high-frequency (UHF) cellular networks \cite{Andrews2011,Zhang2015,Ding2015,Jo2012,Singh2013,Renzo2013,Novlan2013,ElSawy2014,Zhang2017,Onireti2015,Andrew2016}. In the pioneering work on using stochastic geometry for cellular networks \cite{Andrews2011}, it was shown that stochastic model provides a lower bound to real cellular deployment. The work in \cite{Andrews2011} was based on the downlink of cellular networks with the single slope pathloss model. This has been extended  by considering a multi-slope pathloss model in \cite{Zhang2015,Ding2015}, the multi-tier cellular networks in \cite{Jo2012, Singh2013,Renzo2013},   the single tier uplink cellular networks in \cite{Novlan2013} and the multi-tier uplink cellular networks in \cite{ElSawy2014,Zhang2017,Onireti2015}.

The stochastic geometry framework developed for the UHF networks do not directly apply to the mmWave networks due to blockage effects that they suffer from and the much different pathloss model. 
Furthermore, directional beamforming is fundamental in the design of the mmWave cellular networks. 
 The authors in \cite{Bai2015,Renzo2015,Maamari2016,Fang2017,Onireti2017,Onireti2017A} have analyzed the mmWave cellular networks by using the stochastic geometry framework with the blockage effect, realistic pathloss model and the beamforming gain incorporated in their model.  In particular, \cite{Bai2015} pioneered the research work on the  downlink of mmWave cellular  networks by leveraging on an earlier work in  \cite{Bai2014}, which characterized the blockage parameter by some random distribution. Furthermore, the proposed analytical framework in \cite{Bai2015} also captures the significant difference
between the non-line-of-sight (NLOS) and line-of-sight
(LOS) pathloss characteristics. The work in \cite{Bai2015} has been extended to the downlink multi-tier mmWave cellular networks in \cite{Renzo2015}, base station (BS) cooperation in \cite{Maamari2016} and the uplink single-tier mmWave cellular networks in \cite{Fang2017,Onireti2017,Onireti2017A}. 

The uplink analysis for  both the conventional UHF cellular networks and the mmWave cellular networks are deemed to  be quite involved as compared to the downlink analysis due to the per-user power control and the correlation among the interferers \cite{ElSawy2014}. The former is due to the fact that an interfering user could even be closer to a reference mmWave BS than the user that is tagged to the reference mmWave BS. Furthermore, regarding mmWave cellular networks, measurements showed that mmWave
signals propagate with a pathloss exponent of 2 in LOS
paths and a much higher pathloss exponent with additional
shadowing in NLOS paths \cite{Rappaport2013A,Rappaport2013}. 
  This poses a further challenge 
since the difference in pathloss exponents could results in
excessive interference from NLOS users when the per-user power control
is implemented.  Hence, power control must be implemented with a constraint on the maximum user transmit power in order to mitigate the interference. On the other hand, the correlation among the interferer results from the implementation of orthogonal allocation scheme that does not allow for a reuse of a channel resource within the same cell, i.e., the coupling of the mmWave BS and served user-per channel point processes \cite{Novlan2013,ElSawy2014}.
To ensure analytical tractability, various generative models have been proposed in \cite{Novlan2013,ElSawy2014,Zhang2017,Fang2017,Onireti2017,Onireti2017A} to approximate the spatial distribution of interferers in the uplink of UHF and mmWave cellular networks.

In this paper, we present a stochastic geometry framework for modeling and analyzing the uplink in single-tier and multi-tier mmWave cellular networks. Similar to the earlier works in this area \cite{Novlan2013,ElSawy2014,Zhang2017,Fang2017,Onireti2017,Onireti2017A}, we rely on some approximation so as to maintain analytical tractability. Notably, we partially ignore the correlation among the interfering users. 
Our model captures the correlation between the interfering users and the reference mmWave BS, which serves the typical user, but it ignores the  correlation among the interfering users. As evidence from \cite{Novlan2013,ElSawy2014,Zhang2017}, this approximation holds true for the uplink of both the single and multi-tier UHF cellular networks.
 The accuracy of this approach is validated for both the single and multi-tier mmWave cellular networks  via Monte-Carlo simulations. We here extend the work in  \cite{Onireti2017A} which is  based on a single-tier mmWave network and does not take into account the maximum transmit power of the user. Our proposed framework takes into account the limitation in the user transmit power, the per-user power control and the cutoff threshold for the power control.  We compare the findings of our analysis with that from \cite{ElSawy2014} and \cite{Onireti2017A}. The comparison reveals that our analysis provides several new insights that can be leveraged for designing the  mmWave networks more accurately. 
 The main contributions of this work are summarized as follows.
\begin{itemize}
\item 
We present a stochastic geometry framework for the signal-to-interference-and-noise-ratio (SINR) outage probability
in the uplink of a single-tier mmWave cellular networks, which is generic and serving as a foundation extends to the multi-tier mmWave cellular networks. The model takes into account the limitation in the transmit power of the user, and the network defined per-user power control and the cutoff threshold. Closed-form expressions are derived for the SINR outage probability.
\item
We present the asymptotic dense network analysis  of the SINR outage probability in both the single and  multi-tier mmWave cellular networks. The asymptotic  analysis leverage on approximating an intricate LOS function as a step function. 
\item
The analytical derivations are verified via   Monte-Carlo based simulations. Results show that the maximum power constraint significantly affects the SINR outage probability. Further, contrary to the SINR outage of UHF networks, which is non-increasing in  the cutoff threshold, the SINR outage probability in mmWave networks could increase over some range of cutoff thresholds for some mmWave BS density, LOS and NLOS pathloss exponents, and blockage parameter.

 \end{itemize}
The rest of the paper is organized as follows. The system model of the uplink of a multi-tier mmWave cellular network with truncated channel inversion power control is presented in Section \ref{Sec:System}. In Section \ref{Sec_Single}, the uplink modeling framework for a single-tier mmWave cellular network is presented. In Section \ref{Sec:Multi-Tier}, we generalize the developed framework for the multi-tier mmWave cellular networks. In Section \ref{Sec:Dense}, we utilize a simplified system model to analyze the asymptotic behavior and performance in dense mmWave networks.  Numerical and simulation results are presented in Section \ref{Sec:Numerical}. Finally, Section \ref{Sec:Conclusion} concludes the paper.
\vspace{-4mm}
\section{System Model}\label{Sec:System}
\vspace{-2mm}
\subsection{Network Model} 
\vspace{-2mm}
We consider the uplink of a $K$-tier mmWave cellular network and focus on the SINR  experienced by outdoor users served by outdoor mmWave BSs. Each tier's BSs are randomly located and they have particular spatial density,  antenna gain, receiver sensitivity, blockage parameter and pathloss exponents. 
The outdoor BSs of each tier are spatially distributed in $\mathbb{R}^2$ according to an independent homogeneous Poisson point process (PPP) $\Theta_k$ 
 with density $\lambda_k$
. The users locations (before association) are assumed to form a realization of homogeneous PPP $\Phi$
  with density $\lambda_u$. It is assumed that the density of the users is high enough such that each BS will have at least one user served per channel. Each BS serves a single user per channel, which is
randomly selected  from all the users located in its Voronoi cell by using a round-robin scheduler. 
  As in \cite{Novlan2013,ElSawy2014,Andrew2016,Zhang2017,Fang2017,Onireti2017,Onireti2017A}, we assume that the active users also form PPP even after associating just one user per BS. 
  Note that this approximation only partly ignores the correlation imposed by the system model, i.e.,  the coupling of the BS and
served user-per channel resource point processes. The correlation between the reference mmWave BS and the typical user is captured in the derivation of the outage probability in Sections III and IV.
 
  Each tier in the mmWave network is characterized by a non-negative blockage constant $\beta_k$ for $k\in\{1,\ldots,K\}$. The parameter $\beta_k$ is determined by the average size and density of blockages in that tier and where the average LOS range is given by $\frac{1}{\beta_k}$ \cite{Bai2014,Bai2015,Maamari2016}. The probability of a communication link in the $k^{th}$ tier with length $r$ being a LOS is $\mathbb P(\mathrm{LOS}_k)=e^{-\beta_k r}$, while the probability of a link being NLOS is $\mathbb P(\mathrm{NLOS}_k)=1-\mathbb P(\mathrm{LOS}_k)$. The  LOS and NLOS links of the $k^{th}$ tier have different pathloss exponents denoted by $\alpha_L^k $ and $\alpha_N^k$, respectively, $\forall k \in \{1,\ldots, K\}$.

\vspace{-2mm}\subsection{Receiver Sensitivity and Truncated Outage}
We assume that all users have an equal maximum transmit power $P_u$. Furthermore, all mmWave BS in the $k^{th}$ tier have the same receiver sensitivity which is denoted by $\rho_{\min}^k$. The received signal at the mmWave BS must be greater than the receiver sensitivity $\rho_{\min}^k$ for successful transmission in the uplink channel. Hence, each  user (with either LOS or NLOS link to its serving mmWave BS)
  associated with the $k^{th}$ tier 
adjusts its transmit power such that the average  received signal at its serving mmWave BS is equal to a predefined threshold $\rho_o^k$, where $\rho_o^k>\rho_{\min}^k$. Moreover, as a result of the maximum transmit power constraint, 
  users utilize a truncated channel inversion power control scheme, where the transmitters compensate for the pathloss in the link to the receiver to keep the average received signal power to the threshold $\rho_o^k$. Any user-mmWave BS connection that requires a transmit power that exceeds $P_u$ for the pathloss inversion will not be established,  hence, such a connection experiences  a truncation outage \cite{ElSawy2014}.
  


 \vspace{-5mm}
\subsection{Beamforming Gain}
 For analytical tractability, we assume that all users and BSs are equipped with directional antennas with a sectorized gain pattern. The main lobe gain, side lobe gain and beamwidth of the users are $G_u^{\max}$, $G_u^{\min}$ and $\zeta_t$, respectively, while the corresponding parameters of the $k^{th}$ tier BS antennas are  $G_{bk}^{\max}$, $G_{bk}^{\min}$ and $\zeta_{rk}$, respectively. We consider
that based on channel estimation, the reference BS in the $j^{th}$ tier and the
typical user adjust their beam steering angles to achieve the
maximum array gains. As a result of this, the total directivity
gain of the desired signal is $\mathcal{G}_j=G_{bj}^{\max}G_u^{\max}$. Since the underlying PPP is isotropic in $\mathbb{R}^2$, we model the beam directions of the interfering link as a uniform random variable on $[0,2\pi]$. Further, the directivity gain in the interference link $G_l^j$ (interference experienced at the reference BS in the $j^{th}$ tier) can be approximated as discrete random variable whose probability  distribution is given  as $a_v^j$ with probability $b_v^j ~(v\in\{1,2,3,4\})$ \cite{Bai2014}, where $a_v^j$  and $b_v^j$ are defined in Table \ref{Tab_Directivity_Gain}.
\begin{table}
\centering
\caption{Probability Mass Function of the Directivity Gain in an Interference Link of the $j^{th}$ Tier \cite{Bai2014}}
 \vspace{-3mm}
\begin{tabular}{c|c|c|c|c}
\hline 
v & 1 & 2 & 3 & 4 \\ 
\hline 
$a_v^j$ & $G_{bj}^{\max}G_u^{\max}$ & $G_{bj}^{\max}G_u^{\min}$ & $G_{bj}^{\min}G_u^{\max}$ & $ G_{bj}^{\min}G_u^{\min}$ \\ 
\hline 
$b_v^j$ & $\frac{\zeta_{rj}\zeta_t}{4\pi^{2}}$ & $\frac{\zeta_{rj}}{2\pi}(1-\frac{\zeta_t}{2\pi})$ &$(1-\frac{\zeta_{rj}}{2\pi})\frac{\zeta_t}{2\pi}$ & $(1-\frac{\zeta_{rj}}{2\pi})(1-\frac{\zeta_t}{2\pi})$ \\ 
\hline 
\end{tabular}
\label{Tab_Directivity_Gain}
 \vspace{-8mm}
\end{table}

In general, the $k^{th}$ tier is characterized by a set of parameters $\mathcal{V}_k$ whose element include the $k^{th}$ tier's BS density $\lambda_k$, blockage parameter $\beta_k$, cutoff threshold $\rho_o^k$, main lobe gain $G_{bk}^{\max}$, side lobe gain $G_{bk}^{\min}$, beamwidth $\zeta_{rk}$, LOS pathloss exponent $\alpha_L^k$ and its NLOS pathloss exponent $\alpha_N^k$ such that $\mathcal{V}_k=\{\lambda_k,\beta_k,\rho_o^k,G_{bk}^{\max},G_{bk}^{\min}, \zeta_{rk}, \alpha_L^k,\alpha_N^k\}, \forall k=1,\ldots,K$.
 \vspace{-2mm}
\section{Uplink of Single-Tier mmWave Cellular Networks}\label{Sec_Single}
In this section, we develop our framework model for the uplink of a single-tier mmWave cellular network. In particular, we present the mmWave transmission power analysis and the SINR outage probability analysis.
 \vspace{-6mm}
\subsection{mmWave Transmission Power Analysis}
Considering the mmWave cellular network with the  truncated channel inversion scheme, each user, which could be in LOS or NLOS to its serving mmWave BS will transmit with different power in order to invert the pathloss towards its serving  BS. As a result of the truncation channel inversion, not all users will be  able to communicate in the uplink channel\footnote{Note that for all the parameters in  single tier network, we have removed the subscript/superscript $k$ used to distinguish the $k^{th}$-tier parameters  in a multi-tier network.}. 
 In particular, LOS and NLOS users located at a distance greater than $(Pu/\rho_o)^{1/\alpha_L}$ and $(Pu/\rho_o)^{1/\alpha_N}$, respectively, from their associated BS are unable to communicate in the uplink direction as a result of insufficient transmit power \cite{ElSawy2014}. Hence, in addition to the fact that the whole user set is divided into a subset of LOS and NLOS users based on their association with their serving mmWave BS, the LOS and NLOS user sets are further divided into a non-overlapping subset of active users and inactive users. The distribution of the transmit power of a typical user is obtained from the following theorem.
\begin{theorem}\label{lemma:PDF}
In  mmWave cellular networks with truncated channel inversion power control and cutoff threshold $\rho_o$, the probability distribution function (PDF) of the transmit power of a typical user in the uplink is given by
\vspace{-1mm}
\begin{equation}\label{eq:PDF}
f_{P}(p)=\frac{\lambda(p)e^{-\Lambda(p)}}{\int_0^{P_u}\lambda(y)e^{-\Lambda(y)}\mathrm dy},~~~~0\le p \le P_u
\end{equation}
where 
\begin{align}\label{eq:lambda}
\lambda(p) = \frac{2\pi\lambda}{\alpha_L \rho_o^{2/\alpha_L}}p^{\frac{2}{\alpha_L}-1}e^{-\beta\left(\frac{p}{\rho_o}\right)^{\frac{1}{\alpha_L}}}+\frac{2\pi\lambda}{\alpha_N \rho_o^{2/\alpha_N}}p^{\frac{2}{\alpha_N}-1}\left(1-e^{-\beta\left(\frac{p}{\rho_o}\right)^{\frac{1}{\alpha_N}}}\right)
\end{align}
and
\begin{align}\label{eq:Gamma}
\Lambda(p)&=\frac{2\pi\lambda}{\beta^2}\left(1-e^{-\beta\left(\frac{p}{\rho_o}\right)^{\frac{1}{\alpha_L}}}\left(1+\beta\left(\frac{p}{\rho_o}\right)^{\frac{1}{\alpha_L}}\right)\right)+\pi\lambda\left(\frac{p}{\rho_o}\right)^{\frac{2}{\alpha_N}}\\
&-\frac{2\pi\lambda}{\beta^2}\left(1-e^{-\beta\left(\frac{p}{\rho_o}\right)^{\frac{1}{\alpha_N}}}\left(1+\beta\left(\frac{p}{\rho_o}\right)^{\frac{1}{\alpha_N}}\right)\right)\nonumber
.\end{align}

The $\delta^{th}$ moment of the transmit power is thus obtained as
\begin{equation}\label{eq:moment}
\mathbb{E}\left[P^\delta\right]=\int_0^{P_u} p^{\delta}f_{P}(p) \mathrm{d}p,
\end{equation} \label{eq_UE_moment}
where $f_{P}(p)$ is given in (\ref{eq:PDF}). 


  Further, the truncation outage probability, which is the probability  that a user experience outage due to insufficient power, is expressed as
\begin{equation}\label{eq:Trun_Outage1}
\mathcal O_p = e^{-\Lambda(Pu)},
\end{equation}
where ${\Lambda(p)}$ is given in (\ref{eq:Gamma}).  
\begin{proof}
See Appendix \ref{Appen:Proof_PDF}.
\end{proof}
\end{theorem}
 Note that $1-e^{-y}(1+y)$ is strictly increasing in $y$ for $y>0$, hence the first  term of $\Lambda(p)$ in  (\ref{eq:Gamma}) is greater than the third term and $\Lambda(p)$ is strictly positive for all BS density $\lambda$, blockage parameter $\beta$, cutoff threshold $\rho_o$ and pathloss exponent $\alpha_N\ge\alpha_L>0$. Consequently, increasing the cutoff threshold $\rho_o$ leads to increase in the truncation outage probability as long as $\alpha_N\ge\alpha_L$. In other words, the higher the cutoff threshold the poorer the mmWave network performance in terms of the truncation outage. As we will show in the later section, a low cut-off threshold could actually deteriorate the mmWave network  performance in terms of the SINR outage probability. Hence, it is essential to  manage  the trade-off between SINR outage and truncation outage probabilities using the cutoff threshold.
 
  On another note, by expanding (\ref{eq:Gamma}), it can be seen that  increasing the blockage parameter value $\beta$ leads to a reduction in $\Lambda(p)$ for fixed mmWave BS density $\lambda$, cutoff threshold $\rho$, and pathloss exponent $\alpha_N>\alpha_L$. Thus, the truncation outage probability also increases with increasing  blockage parameter $\beta$. Increasing the blockage parameter implies decreasing the average LOS range and hence we have more NLOS paths requiring a much higher transmit power to meet the receiver sensitivity requirement. 

Regarding the expectation of the user transmission power, i.e., the average user transmission power, it is not straightforward to gain insights. However, from (\ref{eq:lambda}), we expect the plot of the average transmit power to be characterized from the LOS-based average transmit power and the NLOS-based average transmit power. We validate this observation later in the numerical results section.
  
 \vspace{-2mm}
\subsection{SINR Outage Probability}
For an active typical user, the SINR at its connected BS (termed as the reference mmWave BS) can be written as
\begin{equation}\label{eq:SINR}
SINR =\frac{\rho_o |g_o|^2G_b^{\max}G_u^{\max}}{\sigma^2+\sum_{z\in\mathcal Z}P_z|g_z|^2G_zL(D_z)},
\end{equation}
where the useful signal power (normalized by $G_b^{\max} G_u^{\max}$) is equal to $\rho_o |g_o|^2$ due to the truncated channel inversion power control, $\mathcal Z$ is the set of interfering users, $L(D_z)$ is the pathloss from the interfering users to the reference mmWave BS,  $\sigma^2$ is the noise power,   $G_z$ is the directivity gain on an interfering link and $g_z$  is the small-scale fading which follows a Nakagami distribution with parameter $N$. The SINR outage probability $\mathcal O_s$ is the probability that the instantaneous SINR experienced at the reference mmWave BS is less than the target SINR $\theta$, i.e. $\mathcal O_s=\mathbb P(SINR<\theta)$. Given that the average received signal at the reference mmWave BS (normalized by the directivity gain $G_b^{\max} G_u^{\max}$) is equal to the cutoff threshold $\rho_o$. The SINR outage probability can be computed as
\begin{equation}
\mathbb{P}(SINR\le\theta)=\mathbb P\{\rho_o|g_o|^2\mathcal{G}\le\theta(\sigma^2+I_L+I_N)\},
\end{equation}
where $I_L$ and $I_N$ are the interference strength from LOS and NLOS users, respectively and $\mathcal{G}=G_b^{\max} G_u^{\max}$. Noting that $|g_o|^2$ is normalized gamma random variable with parameter $N$, we have the following approximation
\begin{align}\label{eq:Prob1}
\mathbb{P}&\{|g_o|^2\le\theta(\sigma^2+I_L+I_N)/(\rho_o\mathcal G)\}\\&\overset{(a)}\approx1-\left(1-\mathbb E\left[\left(1-e^{-\frac{\eta\theta\left(\sigma^2+I_L+I_N\right)}{\rho_o\mathcal G}}\right)^{N}\right]\right)\nonumber\\
&\overset{(b)}=1-\sum_{n=1}^{N}\!\left(-1\right)^{n+1}\binom {N} {n}\mathbb E\left[e^{-\frac{n\eta\theta\left(\sigma^2+I_L+I_N\right)}{\rho_o\mathcal G}}\right]\nonumber\\
&= 1-\sum_{n=1}^{N}\!\left(-1\right)^{n+1}\!\binom {N} {n}\!e^{-sn\sigma^2}\mathcal L_{I_L}(sn)\mathcal L_{I_N}(sn),\nonumber
\end{align}
where $s=\frac{\eta\theta}{\rho_o\mathcal G}$, $\eta=N(N!)^{-\frac{1}{N}}$,   $(a)$ follow from the fact that $\lvert g_{0}\rvert^2$ is a normalized gamma random variable with parameter $N$ and the fact that for a constant $\gamma>0$, the probability $\mathbb P(\lvert g_0\rvert^2<\gamma)$ is tightly upper bounded by $\left[1-\exp\left(-\gamma N\left(N!\right)^{-\frac{1}{N}}\right)\right]^N$ \cite{Alzer1997}. Further, the expectation is with respect to $I_L$ and $I_N$.  $(b)$ follows from the binomial theorem and the assumption that $N$ is an integer, and $\mathcal L_{I_L}$ and $\mathcal L_{I_N}$ denote the Laplace transforms of the random variables $I_L$ and $I_N$, respectively. 

As mentioned earlier, the location of the interfering users do not create a PPP as a result of the correlation among the users from the channel assignment process. The interfering users are thus better modeled by using soft-core processes, which can capture such correlation  \cite{Guo2013}. However, most soft-core processes lack analytical tractability \cite{Haenggi2013,ElSawy2013}, thus making the  expression for the Laplace transforms of the aggregate LOS and NLOS interference, $\mathcal L_{I_L}$ and $\mathcal L_{I_N}$, respectively, unobtainable. Hence, we approximate the location of the interfering users with a PPP. Note that the approximation has been shown to be accurate for the UHF network when the correlation among the interfering nodes and the reference receiver is captured \cite{Novlan2013,ElSawy2014}. Our model  here also captures this correlation. The accuracy of our assumption will be verified later through simulations.
Based on the PPP approximation, and the independent and  identical distributed transmit power for the set of interfering users in the uplink channel, the SINR outage probability can be obtained from the following theorem.

\begin{theorem}\label{Pr:SINR}
The SINR outage probability in the uplink of single-tier mmWave cellular networks with truncated channel inversion power control with cutoff threshold $\rho_o$ can be expressed as
\begin{equation}
\mathcal O_s \!= \!1-\sum_{n=1}^{N}\left(-1\right)^{n+1}\binom {N} {n}\exp\left(-\frac{\eta n\theta\sigma^2}{\rho\mathcal G}-Q_n-V_n\right)
\end{equation}
where
\begin{align}
Q_n=2\pi\lambda
\sum_{v=1}^{4}\!b_vq_v^{\frac{2}{\alpha_L}}\int_{\mathcal{A}}^\infty \int_{0}^{P_u}\mathcal F\left(N,\frac{y^{-\alpha_L}}{N}\right)\exp\left({-\beta \left(q_vp\right)^{\frac{1}{\alpha_L}}y}\right)yP^{\frac{2}{\alpha_L}}f_P(p)\mathrm dp\mathrm dy,
\end{align}
\begin{align}
V_n=2\pi\lambda\sum_{v=1}^{4}\!b_vq_v^{\frac{2}{\alpha_N}}\int_{\mathcal{B}}^\infty\int_{0}^{P_u}\mathcal F\left(N,\frac{y^{-\alpha_N}}{N}\right)\left(1-\exp\left({-\beta \left(q_vp\right)^{\frac{1}{\alpha_N}}y}\right)\right)yP^{\frac{2}{\alpha_N}}f_P(p)\mathrm dp\mathrm dy,
\end{align}
where   $\eta=N(N!)^{-\frac{1}{N}}$, $\mathcal G=G_u^{\max}G_b^{\max}$, $\mathcal{A}=\left(\frac{\eta n\theta a_v}{\mathcal G}\right)^{-\frac{1}{\alpha_L}}$, $\mathcal B=\left(\frac{\eta n\theta a_v}{\mathcal G}\right)^{-\frac{1}{\alpha_N}}$,  $q_k=\frac{\eta n\theta a_v}{\rho_o\mathcal G}$,   $\mathcal F(N,y)=1-\frac{1}{\left(1+y\right)^N}$, $a_v$ and $b_v$ are the antenna directivity parameters defined in Section \ref{Sec:System} and $f_P(p)$ is defined in (\ref{eq:PDF})
\begin{proof}
See Appendix \ref{Appen:Proof_SINR}
\end{proof}
\end{theorem} 
Though this approximates the SINR outage probability, we find that the expression compares very well with the simulation results in Section \ref{Sec:Num_Accu}. Furthermore, the expression here captures the user maximum power constraint contrary to the prior result on the uplink of  single tier mmWave networks in \cite{Onireti2017A},  which is based on an unbounded power constraint.  The maximum power constraint is very important in the uplink power control of mmWave network due to the significant difference in the LOS and NLOS pathloss exponent. We show the impact of the maximum power constraint in a single-tier network later in Section \ref{Num:Sect_Comp_prior}.
%
 \vspace{-2mm}
\section{Uplink of  Multi-Tier mmWave Cellular Networks}\label{Sec:Multi-Tier}
In this section, we extend our developed model for the uplink of a single-tier mmWave cellular network to the uplink of a multi-tier mmWave cellular network. As mentioned earlier, the $k^{th}$ tier is characterized by a set 
 $\mathcal{V}_k=\{\lambda_k,\beta_k,\rho_o^k,G_{bk}^{\max},G_{bk}^{\min}, \zeta_{rk}, \alpha_L^k,\alpha_N^k\}, \forall k=1,\ldots,K$. First, we present the distribution of transmit power for the multi-tier mmWave cellular networks. Afterward, we derive its SINR outage probability.
 \vspace{-2mm}
\subsection{Distribution of the Transmit Power in the Uplink of Multi-tier mmWave Cellular Networks}
Similar to the single-tier network, given the cutoff threshold for the $k^{th}$ tier $\rho_o^k$, LOS and NLOS users located at distances greater than $(P_u/\rho_o^k)^{1/\alpha_L^k}$ and $(P_u/\rho_o^k)^{1/\alpha_N^k}$, respectively, from their nearest mmWave BS are unable to communicate in the uplink direction due to insufficient transmit power. The distribution of the transmit power of a typical user associated with the $j^{th}$ tier is obtained from the following theorem
\begin{theorem}\label{Lemma:PDF_Multi}
 In a $K$-tier mmWave cellular network with truncated channel inversion power control where the $k^{th}$ tier is distinguished by the set $\mathcal V_k$, $\forall k=1,\ldots,K$, i.e., its density $\lambda_k$, blockage parameter $\beta_k$, cutoff threshold $\rho_o^k$, antenna parameters, $G_{bk}^{\max}$, $G_{bk}^{\min}$ and $\zeta_{rk}$, LOS pathloss exponent  $\alpha_L^k$ and its NLOS pathloss exponent $\alpha_N^k$, the PDF of the transmit power of a typical active user in the uplink  of the $j^{th}$ tier is given by
\begin{equation}\label{eq:PDF_multi}
f_{P_j}(p)=\frac{\sum_{k=1}^{K}\overline{\lambda}_k(p)}{1-e^{-\sum_{a=1}^K{ \Lambda_a\left(\frac{P_u}{\rho_o^j}\right)}}}{e^{-\sum_{b=1}^K{ \Lambda_b\left(\frac{p}{\rho_o^j}\right)}}}
\end{equation}
where
\begin{equation}
\overline{\lambda}_k(p) =\frac{2\pi\lambda_k}{\alpha_L^k {\rho_o^j}^{2/\alpha_L^k}}p^{\frac{2}{\alpha_L^k}-1}e^{-\beta_k\left(\frac{p}{\rho_o^j}\right)^{\frac{1}{\alpha_L^k}}}+\frac{2\pi\lambda_k}{\alpha_N^k {\rho_o^j}^{2/\alpha_N^k}}p^{\frac{2}{\alpha_N^k}-1}\left(1-e^{-\beta_k\left(\frac{p}{\rho_o^j}\right)^{\frac{1}{\alpha_N^k}}}\right),
\end{equation}
\begin{equation}\label{eq:Lambda_multi_tier}
\Lambda_k(y)=\frac{2\pi\lambda_k}{\beta_k^2}\left(1-e^{-\beta_k{y}^{\frac{1}{\alpha_L^k}}}\left(1+\beta_k{y}^{\frac{1}{\alpha_L^k}}\right)\right)+\pi\lambda_k{y}^{\frac{2}{\alpha_N^k}}
-\frac{2\pi\lambda_k}{\beta_k^2}\left(1-e^{-\beta_k{y}^{\frac{1}{\alpha_N^k}}}\left(1+\beta_k{y}^{\frac{1}{\alpha_N^k}}\right)\right)
\end{equation}
and $y$ is a dummy variable in (\ref{eq:Lambda_multi_tier}).
The $\delta^{th}$ moment of the transmit power of a user in the $j^{th}$ tier is given as
\begin{equation}
\mathbb[P_j^\delta]=\int_0^{P_u}{\frac{p^{\delta}\sum_{k=1}^{K}\overline{\lambda}_k(p)}{1-e^{-\sum_{a=1}^K{ \Lambda_a\left(\frac{P_u}{\rho_o^j}\right)}}}{e^{-\sum_{b=1}^K{ \Lambda_b\left(\frac{p}{\rho_o^j}\right)}}}}\mathrm{d}p
\end{equation}
Further, the truncation outage probability in the uplink of mmWave cellular networks for the $j^{th}$ tier can be obtained as
\begin{equation}
\mathcal O_p^j=e^{-\sum_{k=1}^K{ \Lambda_k\left(\frac{P_u}{\rho_o^j}\right)}}.
\end{equation}
\begin{proof}
See Appendix \ref{Appen:Proof_PDF_Multi}
\end{proof}
\end{theorem}

\subsection{SINR Outage Probability}
For an active typical user, the SINR at its connected BS in the $j^{th}$ tier (termed as the reference mmWave BS) can be written as
\begin{equation}\label{eq:SINR_M}
SINR_j =\frac{\rho_o^j |g_o|^2\mathcal G_{j}}{\sigma^2+\sum_{k=1}^K\sum_{z\in\mathcal Z_k}P_{z}|g_{z}|^2G_z^jL(D_z)},
\end{equation}
where  the useful signal power (normalized by $\mathcal G_{j}$) is equal to $\rho_o^j |g_o|^2$ due to the truncated channel inversion power control, $\mathcal Z_k$ is the set of interfering users associated with a BS in the $k^{th}$ tier, $L(D_z)$ is the pathloss from the interfering users to the reference BS,  $\sigma^2$ is the noise power,   $G_z^j$ is the directivity gain on an interfering link and $g_z$  is the small-scale fading which follows a Nakagami distribution with parameter $N$. We consider that each of the tiers has its own SINR threshold which is represented by $\theta_k$. Further, the average received signal at any of the BSs in the $j^{th}$ tier is equivalent to the cutoff threshold  of the $j^{th}$ tier represented by $\rho_o^j$.

The SINR outage probability of the $j^{th}$ tier can be expressed as
\begin{equation}
\mathbb{P}(SINR_j\le\theta_j)=\mathbb P\left\{\rho_o^j|g_o|^2\mathcal G_j\le \theta_j\left(\sigma^2+\sum_{k=1}^KI_L^k+\sum_{k=1}^KI_N^k\right)\right\},
\end{equation}
where $I_L^k$ and $I_N^k$ are the aggregate interference from LOS and NLOS users of the $k^{th}$ tier, respectively. Note that $I_L^j$ and $I_N^j$ represents the LOS and NLOS co-tier interference, respectively, and  $I_L^k$ and $I_N^k$ $\forall k\neq j$ denotes the LOS and NLOS cross-tier interference, respectively. Similar to the single tier case, noting that $|g_o|^2$ is normalized gamma random variable with parameter $N$, we have the following approximation
\begin{align}
\mathbb{P}&\left\{|g_o|^2\le\theta_j\left(\sigma^2+\sum_{k=1}^KI_L^k+\sum_{k=1}^KI_N^k\right)/\left(\rho_o^j\mathcal G_j\right)\right\}\\
&= 1-\left(1-\mathbb E\left[\left(1-e^{-\frac{\eta\theta_j\left(\sigma^2+\sum_{k=1}^KI_L^k+\sum_{k=1}^KI_N^k\right)}{\rho_o^j\mathcal G_j}}\right)^N\right]\right)\nonumber\\
&=1-\sum_{n=1}^N(-1)^{n+1}\binom{N}{n}\mathbb E_\Phi\left[e^{-\frac{n\eta\theta_j\left(\sigma^2+\sum_{k=1}^KI_L^k+\sum_{k=1}^KI_N^k\right)}{\rho_o^j\mathcal G_j}}\right]\nonumber\\
&=1-\sum_{n=1}^N(-1)^{n+1}\binom{N}{n}e^{-sn\sigma^2}\prod_{k=1}^K\mathcal L_{I_L^k}(sn)\prod_{k=1}^K\mathcal L_{I_N^k}(sn),\nonumber
\end{align}
where $s=\frac{\eta\theta_j}{\rho_o^j\mathcal G_j}$, $\eta=N(N!)^{-\frac{1}{N}}$. In addition, the interfering users do not constitute a PPP while the transmit power of the interfering users are also correlated as in the case of the single tier. However, in order to keep analytical tractability, we approximate the interfering users process as a PPP while also ignoring the correlations in the transmit power of interfering users. The SINR outage probability of a typical user in the $j^{th}$ tier of mmWave cellular networks can be obtained from the following theorem.
\begin{theorem} \label{Theorem:SINR_Multi}
In a $K-$tier mmWave network with truncated channel inversion power control where each tier is distinguished by its density $\lambda_k$, cutoff threshold $\rho_o^k$, blockage parameter $\beta_k$, antenna parameters, $G_{bk}^{\max}$, $G_{bk}^{\min}$ and $\zeta_{rk}$,  LOS pathloss exponent $\alpha_L^k$ and NLOS pathloss exponent $\alpha_N^k$, the SINR outage probability of a typical user in the $j^{th}$ tier is given by

\begin{equation}
\mathcal O_s^j = 1-\sum_{n=1}^{N}\left(-1\right)^{n+1}\binom {N} {n}\exp\left(-\frac{\eta n\theta_j\sigma^2}{\rho_o^j\mathcal G_j}-\sum_{k=1}^K\left(Q_n^k+V_n^k\right)\right)
\nonumber\end{equation}
where

\begin{align}
Q_n^k=2\pi\lambda_k\sum_{v=1}^{4}b_v^j{q_v^j}^{\frac{2}{\alpha_L^j}}\nonumber\times\int_{\mathcal{A}_j}^\infty \int_{0}^{P_u}\mathcal F\left(N,\frac{y^{-\alpha_L^j}}{N}\right)e^{-\beta_j \left(q_v^jp\right)^{\frac{1}{\alpha_L^j}}y}yp^{\frac{2}{\alpha_L^j}}f_{P_k}(p)\mathrm dp\mathrm dy,
\nonumber
\end{align}

\begin{align}
V_n^k= 2\pi\lambda_k\sum_{v=1}^{4}\!b_v^j{q_v^j}^{\frac{2}{\alpha_N^j}}\times\nonumber\int_{\mathcal{B}_j}^\infty \int_{0}^{P_u}\mathcal F\left(N,\frac{y^{-\alpha_N^j}}{N}\right)\left(1-e^{-\beta_j \left(q_v^jp\right)^{\frac{1}{\alpha_N^j}}y}\right)yp^{\frac{2}{\alpha_N^j}}f_{P_k}(p)\mathrm dp\mathrm dy,
\nonumber\end{align}
 $\eta=N(N!)^{-\frac{1}{N}}$, 
  $\mathcal{A}_j=\left(\frac{\eta n\theta_ja_v^j\rho_o^k}{\rho_o^j\mathcal G_{j}}\right)^{-\frac{1}{\alpha_L^j}}$, $\mathcal B_j=\left(\frac{\eta n\theta_ja_v^j \rho_o^k}{\rho_o^j\mathcal G_{j}}\right)^{-\frac{1}{\alpha_N^j}}$,   $\mathcal F(N,y)=1-\frac{1}{\left(1+y\right)^N}$, $q_v^j=\frac{\eta n\theta_j a_v^j}{\rho_o^j\mathcal G_{j}}$, $a_v^j$ and $b_v^j$ are the antenna directivity parameters defined in Section \ref{Sec:System} and $f_{P_k}(p)$ is defined in (\ref{eq:PDF_multi}).


\begin{proof} 
See Appendix \ref{Appen:Proof of SINR_Multi}
\end{proof}
\end{theorem}
It has been shown in \cite{Bai2015} that the LOS probability function can be approximated by a step function in a dense mmWave network. Hence, in the next section, we propose to simplify our uplink  system model and the subsequent analysis by using a step function approximation of the LOS probability function as well. 
 \vspace{-2mm}
\section{Analysis of the Uplink of Dense mmWave Networks}\label{Sec:Dense}
In this section, we present the analysis for the uplink of a dense mmWave cellular network. The motivation for the dense network analysis is based on the fact that mmWave cellular networks must be dense in order to achieve its forecasted gain \cite{Bai2015}. Here we approximate the LOS probability $\mathbb P(\mathrm{LOS}_k)$ by using a step function such that the LOS probability $\mathbb P(\mathrm{LOS}_k)$ is taken to be $1$ when the link is within a circular disc $\mathcal B(0,R_B)$ centered at the reference mmWave BS  and $0$ when outside the disc. 
Next, we present  uplink signal to interference ratio (SIR) distribution for the dense multi-tier  mmWave networks, which we later degrade to the single tier scenario.

  \vspace{-2mm}
 \subsection{Outage Analysis in the Uplink of Dense Multi-tier mmWave Networks}
 The dense mmWave network will be  interference limited with mainly LOS interferers limiting its performance. Hence, we ignore both the noise power and the NLOS interfering users in the analysis. Further,
  according to \cite{Rappaport2013}, the signal power from LOS interferers are nearly deterministic, hence we also ignore the small-scale fading.
 Consequently, the  SIR at the BS that the typical user
connects to in the $j^{th}$ tier can be expressed from (\ref{eq:SINR_M}) as
 \begin{equation} 
  SIR_j=\frac{\rho_o^j\mathcal G_j}{\sum_{k=1}^K\sum_{z\in\mathcal Z_k\cap\mathcal B(0,R_B)}P_zG_z^jL(D_z)}.
 \end{equation}
 As mentioned earlier, for the dense deployment, the LOS interferers are dominant and the SINR outage probability in the $j$ tier can thus be approximated as
 \begin{equation}
 \mathbb P(SIR_j\le\theta_j)=\mathbb P\left\{\rho_o^j\mathcal G_j\le\theta_j\sum_{k=1}^K I_L^k\right\},
 \end{equation} 
where $I_L^k=\sum_{z\in \mathcal Z_k\cap \mathcal B(0,R_B)}P_zG_zL(D_z)$ is the interference power received at the reference BS from users connected to BSs in the $k^{th}$ tier. Note that  the average receive signal at the reference BS normalized by the directivity gain $\mathcal G_j$ is equivalent to the cutoff threshold $\rho_o^j$. The SINR outage probability can be approximated as
\begin{align}\label{eq:outage_dense1}
\mathbb P\{\rho_o^j\mathcal G_j<\theta_j\sum_{k=1}^K I_L^k\}&\overset{(a)}\approx\mathbb{P}\left\{h<\frac{\theta_j \sum_{k=1}^KI_L^k}{\rho_o^j\mathcal G_j}\right\}\nonumber\\
&\overset{(b)}=1-\left(1-\mathbb{E}_{\Phi_L}\left[\left(1-e^{-\frac{\eta \theta_j \sum_{k=1}^KI_L^k}{\rho_o^j\mathcal G_j}}\right)^L\right]\right)\nonumber\\
&=1-\sum_{l=1}^L
\binom{L}{l}(-1)^{l+1}\prod_{k=1}^K \mathcal{L}_{I_L^k}(sl)
\end{align} 
where the dummy variable $h$ in $(a)$ is used to denote normalized gamma variable with parameter $L$. Note that the distribution of the normalized gamma variable converges to an identity when its parameters tend to infinity, $(b)$ follows from \cite{Alzer1997} such that  the probability $\mathbb P(\lvert h\rvert^2<\gamma)$ is tightly upper bounded by $\left[1-\exp\left(-\gamma N\left(N!\right)^{-\frac{1}{N}}\right)\right]^N$ and   $s=\frac{\eta \theta_j}{\rho_o^j\mathcal G_j}$.
  In the following theorem, we summarize the main result for the SIR outage distribution in a multi-tier network.
 \begin{theorem}\label{Pr:SINR_ultra}
 The SINR outage probability in the $j^{th}$ tier of a $K$-tier mmWave cellular network with truncated channel inversion power control can be approximated as
 \begin{align}\label{eq:SINRoutage_dense}
 \overline{\mathcal O}_s^j&=1-\sum_{n=l}^L(-1)^{l+1}\binom{L}{l}\prod_{k=1}^K\exp\Bigg(\int_0^{Pu}\bigg(\pi\lambda_k\left(\frac{p}{\rho_o^k}\right)^{\frac{2}{\alpha_L^j}}-\lambda_0^k\\&+\frac{2\pi\lambda_k}{\alpha_L^j}\sum_{v=1}^4\left(\frac{\eta l\theta_j a_v^j}{\rho_o^j\mathcal G_j}\right)^{\frac{2}{\alpha_L^j}}b_v^jp^{\frac{2}{\alpha_L^j}}\left(\Gamma\left(\frac{-2}{\alpha_L^j},\frac{\eta l\theta_j a_v^jp}{\rho_o^j\mathcal G_jR_B^{\alpha_L}}\right)-\Gamma\left(\frac{-2}{\alpha_L^j},\frac{\eta l\theta_j a_v^j\rho_o^k}{\rho_o^j\mathcal G_j}\right)\right)\bigg)f_{P_k}(p)\mathrm dx\Bigg)
 \nonumber
 \end{align}
 where $\lambda_0^k=\lambda_k\pi  R_B^2$, $\Gamma(a,b)=\int_b^\infty t^{a-1}e^{-t}\mathrm dt$ is the upper incomplete gamma function, $a_v^j$ and $b_v^j$ are the antenna directivity parameters defined in Section II, $\eta=L(L!)^{-\frac{1}{L}}$, $L$ is used in the approximation and $f_{P_k}(p)$ is defined as
  \begin{equation}
  f_{P_k}(p)=\frac{\sum_{c=1}^K\frac{2\pi\lambda_c p^{\frac{2}{\alpha_L^c}-1}}{\alpha_L^c\left(\rho_o^k\right)^{\frac{2}{\alpha_L^c}}}}{1-e^{-\sum_{a=1}^K\pi\lambda_a\left(\frac{P_u}{\rho_o^k}\right)^{\frac{2}{\alpha_L^a}}}}e^{-\sum_{b=1}^K\pi\lambda_b\left(\frac{p}{\rho_o^k}\right)^{\frac{2}{\alpha_L^b}}}
 \end{equation}

 \begin{proof}
 See Appendix \ref{Appen:Proof_SINR_Ultra}. 
 \end{proof}

 \end{theorem}

  For a single-tier dense mmWave network, the approximation of SINR outage probability can be obtained from the following corollary.
  
 \begin{corollary}\label{Pr:SINR_ultra_Multi_Tier}
 The SINR outage probability in the $j^{th}$ tier of a $K$-tier mmWave cellular network with truncated channel inversion power control can be approximated as
\begin{align}
 \overline{\mathcal O}_s=1-&\sum_{n=l}^L(-1)^{l+1}\binom{L}{l}\exp\Bigg(\int_0^{Pu}\bigg(\pi\lambda\left(\frac{p}{\rho_o}\right)^{\frac{2}{\alpha_L}}-\lambda_0\\&+\frac{2\pi\lambda}{\alpha_L}\sum_{v=1}^4\left(\frac{\eta l\theta a_v}{\rho_o\mathcal G}\right)^{\frac{2}{\alpha_L}}b_vp^{\frac{2}{\alpha_L}}\left(\Gamma\left(\frac{-2}{\alpha_L},\frac{\eta l\theta a_vp}{\rho_o\mathcal GR_B^{\alpha_L}}\right)-\Gamma\left(\frac{-2}{\alpha_L},\frac{\eta l\theta a_v}{\mathcal G}\right)\right)\bigg)f_P(p)\mathrm dp\Bigg)
 \nonumber\end{align}
 where $\lambda_0=\lambda\pi  R_B^2$, $\Gamma(a,b)=\int_b^\infty t^{a-1}e^{-t}\mathrm dt$ is the upper incomplete gamma function, $a_v$ and $b_v$ are the antenna directivity parameters defined in Section II, $\eta=L(L!)^{-\frac{1}{L}}$, $L$ is the number of terms used in the approximation and $f_P(p)$ is defined as
 \begin{equation}
 f_P(p)=\frac{2\pi\lambda p^{\frac{2}{\alpha_L}-1}e^{-\pi\lambda\left(\frac{p}{\rho_o}\right)^{\frac{2}{\alpha_L}}}}{\alpha_L\rho_o^{\frac{2}{\alpha_L}}\left(1-e^{-\pi\lambda\left(\frac{P_u}{\rho_o}\right)^{\frac{2}{\alpha_L}}}\right)}.
 \end{equation}
 \begin{proof}
 The proof follows directly from proof of the multi-tier mmWave cellular networks and is omitted here.
 \end{proof}
 \end{corollary}
 
\vspace{-2mm}
\section{Numerical Results}\label{Sec:Numerical}

In this section, we present numerical results to illustrate our analytical findings for both the single-tier and two-tier mmWave cellular networks. Unless otherwise stated, we set the BS densities $\lambda_1= 10~\mathrm{BS/km^2}$ and $\lambda_2=2\lambda_1$, the maximum transmit power $P_u=1~\mathrm{W}$, $\sigma^2=-110~\mathrm{dBm}$, the tier blockage parameters $\beta_1=0.0071$ and $\beta_2=0.0143$ with corresponding pathloss exponent $\alpha_L^1=2,\alpha_N^1=4, \alpha_L^2=2.9,\alpha_N^2=5$, the Nakagami fading parameter $N=3$. Further, unless otherwise stated,  the antenna parameters of the first and second tier BSs are equivalent such that  $G_{b1}^{\max}=G_{b2}^{\max}=\mathrm{7~dB},~G_{b1}^{\min}=G_{b2}^{\max}=\mathrm{-10~dB}$ and $\zeta_{b1}=\zeta_{b2}=30^{\circ}$, while that of the users are assumed to be characterized with $G_u^{\max}=\mathrm{7~dB,}~G_u^{\min}=\mathrm{-10~dB}$ and $\zeta_u=90^{\circ}$. In addition,  we have utilized the system parameters of the first tier for the single-tier network results.
\subsubsection{Accuracy of Analysis}\label{Sec:Num_Accu}
\begin{figure}
 \begin{center}  
 \includegraphics[width = .5\textwidth]{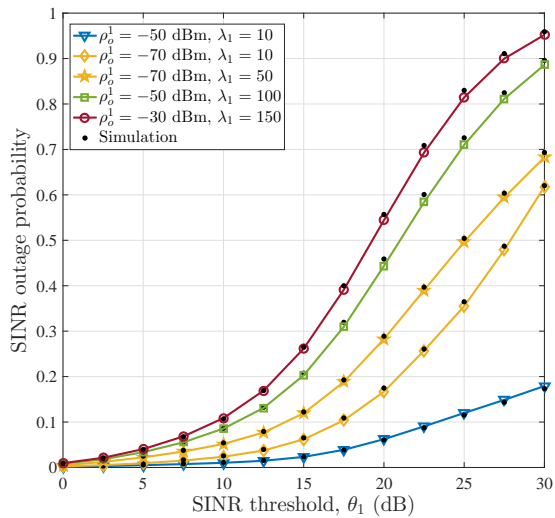} 
 \vspace{-4mm}
  \caption{\small Comparison of the analytical results with simulation in  single-tier mmWave newtorks for $\beta_1=0.0071$, $\alpha_L^1 = 2$ and $\alpha_N^1 = 4$.  \label{fig:SINR_Threshold_Simulation_verification}}
   \end{center} 
   \vspace{-10mm}
     \end{figure}
     
     \begin{figure}
 \begin{center}  
 \includegraphics[width = .5\textwidth]{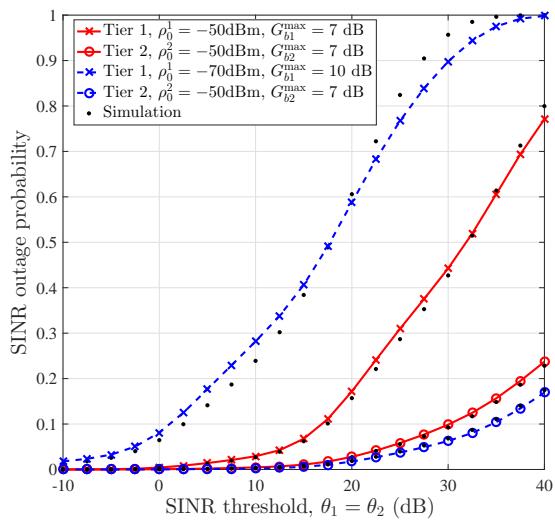} 
 \vspace{-4mm}
  \caption{\small Comparison of the analytical results with simulation in a two-tier mmWave newtork for $\beta_1=0.0071, \beta_2=0.0143$, $\alpha_L^1 = 2$, $\alpha_N^1 = 4, \alpha_L^2=2.9, \alpha_N^2=5$ and BS densities $\lambda_1=10\mathrm{BS/km^2}$ and  $\lambda_2=2\lambda_1$. Red  and blue dashed-lines represent  mmWave networks 1 and 2, respectively. \label{fig:SINR_Threshold_Simulation_verification_2_tier}}
   \end{center} 
   \vspace{-10mm}
     \end{figure}
In Figs. \ref{fig:SINR_Threshold_Simulation_verification} and  \ref{fig:SINR_Threshold_Simulation_verification_2_tier}, we verify our derivation by plotting the analytical and simulation results for the single-tier and two-tier mmWave cellular networks, respectively. The results show that our derived analytical model accurately captures the SINR outage probability for both the single-tier and multi-tier mmWave cellular networks. Hence, our derived model finds great application in mmWave multi-tier network where each tier can be  identified via its BS density, blockage parameter and corresponding LOS and NLOS  pathloss exponents, receiver sensitivity, and the BS antenna gain. Note that this validation is essential since the cumulative distribution of the SINR is based on the assumption that the active user constitute a PPP and that the transmit powers of the users are independent. Further, independent LOS  probability was assumed. The approach in this paper, however, captures the correlation between the location of the reference mmWave BS and that  of the interfering users. It also captures the correlation between  the typical user's (served by the reference mmWave BS)  transmit power  and the interfering users' transmit powers.


\subsubsection{Comparison with the case without maximum power constraint} \label{Num:Sect_Comp_prior}
\begin{figure}
 \begin{center}  
 \includegraphics[width = .5\textwidth]{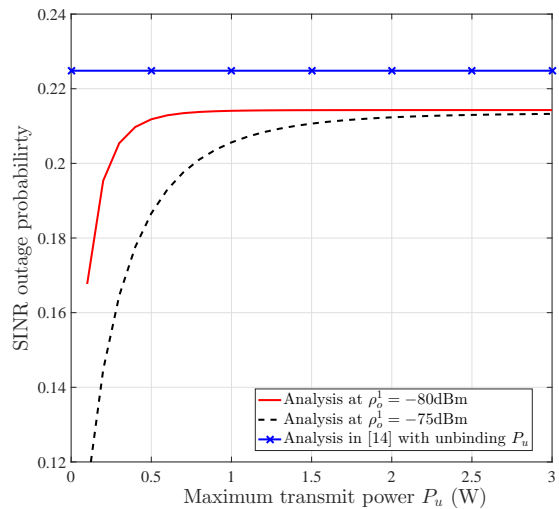} 
 \vspace{-4mm}
  \caption{\small Effect of the maximum transmit power in a single-tier mmWave newtork for $\theta_1=20~\mathrm{dB}, \beta_1=0.0071$, $\alpha_L^1 = 2$, $\alpha_N^1 = 4$ and BS densities $\lambda=10~\mathrm{BS/km^2}$.  \label{fig:SINR_outage_eff_Max_power_lam_10}}
   \end{center} 
   \vspace{-8mm}
     \end{figure}

In Fig. \ref{fig:SINR_outage_eff_Max_power_lam_10}, we compare our analysis with the one presented in \cite{Onireti2017A}, which  does not incorporate the maximum user power constraint. Note that the analysis in \cite{Onireti2017A} is for the single-tier, and hence the comparison presented in Fig. \ref{fig:SINR_outage_eff_Max_power_lam_10} is also based on a single tier. As it can be seen, the maximum power constraint significantly affects the SINR outage probability.  The figure shows that the SINR outage probability derived in \cite{Onireti2017A} does not vary with $P_u$, since the maximum power constraint is ignored in \cite{Onireti2017A}.  

\subsubsection{Effect of the BS Density}
\begin{figure}
 \begin{center}  
 \includegraphics[width = .5\textwidth]{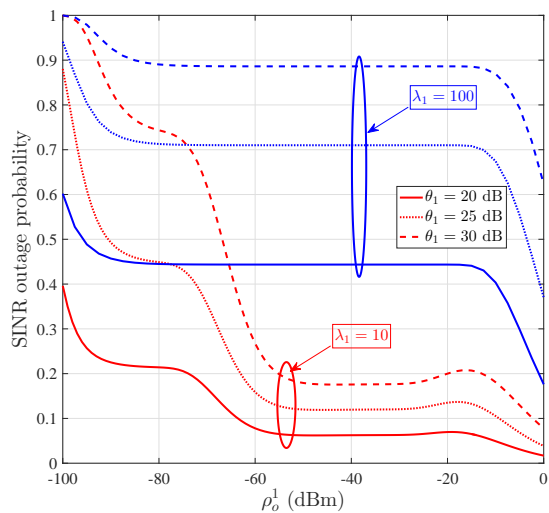} 
  \vspace{-4mm}
  \caption{\small Effect of the BS density on the SINR outage probability   for  
 $\lambda_1=10,100~\mathrm{BS/km^2}$, $\beta_1=0.0071$, $\alpha_L^1 = 2$ and $\alpha_N^1 = 4$ in a single tier network . \label{fig:SINR_outage_eff_Threshold_lambda}}
   \end{center} 
   \vspace{-10mm} 
     \end{figure}
     In Fig. \ref{fig:SINR_outage_eff_Threshold_lambda}, we plot the SINR outage probability for the uplink of a single-tier mmWave networks with truncated channel inversion  power control for SINR threshold $\theta = 20,25$ and $30 \mathrm{dB}$, and BS densities $\lambda_1=10, 100~\mathrm{BS/km^2}$.  It can be seen that the SINR outage probability of mmWave deviates from that of the UHF network presented in \cite{ElSawy2014}. 
More specifically, four sections can be identified from the plot for the BS density $\lambda_1=10~\mathrm{BS/km^2}$: 1) a decrease in SINR outage probability can be seen for the cutoff threshold $\rho_o^1$ ranging from $-100$ to $-50$ dBm with a slow descent region observed for $\rho_0^1$ ranging from $-85$ to $-75$ dBm; 2) a fairly stable outage probability can be observed for $\rho_o^1$ ranging from $-50$ to $-31$ dBm; 3) an increase in SINR outage probability can be seen for   $\rho_o^1$ ranging from $-31$ to $-18$ dBm, and 4) a decrease in SINR outage probability can be seen for $\rho_o^1$ ranging from $-18$ to $0$ dBm.
This observation  is as a result of the large difference in the pathloss exponent of the LOS and NLOS propagation path, with each having its dominance region which also depends on the BS density and blockage parameter. The latter specifies the LOS range. Further, the receiver sensitivity  also specifies the density of active LOS and NLOS users and consequently, the interference received at the reference BS.  It can also be observed from Fig. \ref{fig:SINR_outage_eff_Threshold_lambda}  that   for the same SINR threshold,  increasing the BS density leads to an increase in the SINR outage probability. 

  \subsubsection{Effect of the Blockage}
  \begin{figure}
 \begin{center}  
 \includegraphics[width = .5\textwidth]{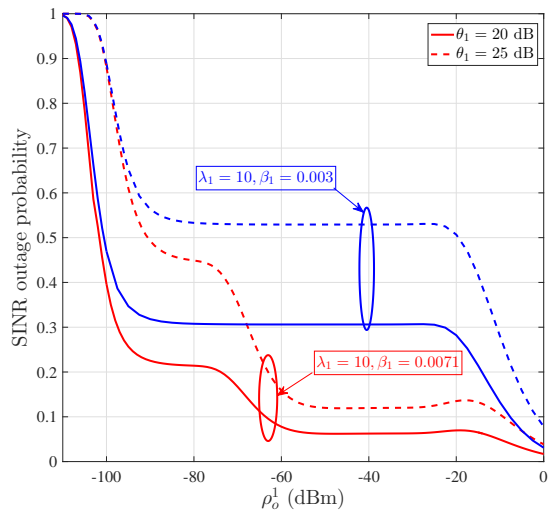} 
  \vspace{-4mm}
  \caption{\small Effect of the blockage parameter on the SINR outage probability   for $\beta_1= 0.0071,0.003$, $\lambda_1=10~\mathrm{BS/km^2}$,  $\alpha_L^1 = 2$ and $\alpha_N^1 = 4$  in a single tier network. 
  \label{fig:SINR_outage_eff_Threshold_beta}}
   \end{center} 
   \vspace{-10mm} 
     \end{figure}In Fig. \ref{fig:SINR_outage_eff_Threshold_beta}, we show the effect of blockages on the SINR outage probability. Based on the LOS probability function $e^{-\beta_1 r}$, a lower $\beta_1$  yields a larger number of LOS interfering user. Hence, the interference power increases when $\beta_1$ is lowered leading to a higher SINR outage probability  for a lower $\beta_1$, as it can be seen in  Fig. \ref{fig:SINR_outage_eff_Threshold_beta}.
   

\subsubsection{Truncation Outage Probability}
\begin{figure}
 \begin{center}  
 \includegraphics[width = .5\textwidth]{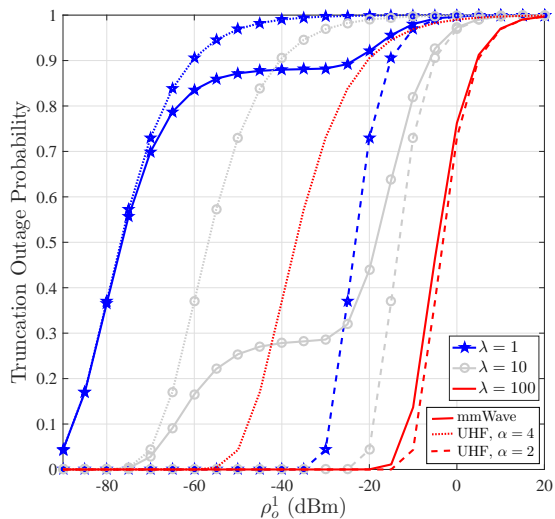} 
 \vspace{-4mm}
  \caption{\small Comparison of the truncation outage probability of mmWave and UHF cellular networks for BS density $\lambda=1,10$ and $100~\mathrm{BS/km^2}$. $\beta_1= 0.0071$, $\alpha_L^1=2$ and $\alpha_N^1=4$ in the mmWave network.
   \label{fig:Trunc_out_beta1_140}}
   \end{center} 
   \vspace{-10mm} 
     \end{figure}

 Fig. \ref{fig:Trunc_out_beta1_140} compares the truncation outage probability for the uplink of mmWave and UHF cellular networks for BS density $\lambda=1,10$ and $100~\mathrm{BS/km^2}$. The truncation outage probability  of UHF networks has been defined in \cite{ElSawy2014}.  It can be seen that similar to the UHF case, increasing the cutoff threshold increases the outage probability since more users are unable to communicate due to insufficient transmit power. These results are in line with the insights previously drawn on equations (\ref{eq:Gamma}) and (\ref{eq:Trun_Outage1}).  Furthermore, for BS densities $\lambda=1,10$, the truncation outage of mmWave networks experience a slow growth region as the cutoff threshold increases before its saturation contrary to UHF networks, which does not experience a slow growth region. The slow growth region is due to the difference in the truncation outage probability for LOS and NLOS links at a given cutoff threshold. Meanwhile, for a high BS density of $\lambda=100$, the truncation outage probability of mmWave converges to that of UHF with $\alpha=2$ since more paths becomes LOS as the BS density increases. 
  As expected, Fig. \ref{fig:Trunc_out_beta1_140} shows that the truncation outage of mmWave networks reduces with as the BS density increases. This observation is due to the  shortening of the average link lengths as the BS density is increased.  
     
\begin{figure}
 \begin{center}  
 \includegraphics[width = .5\textwidth]{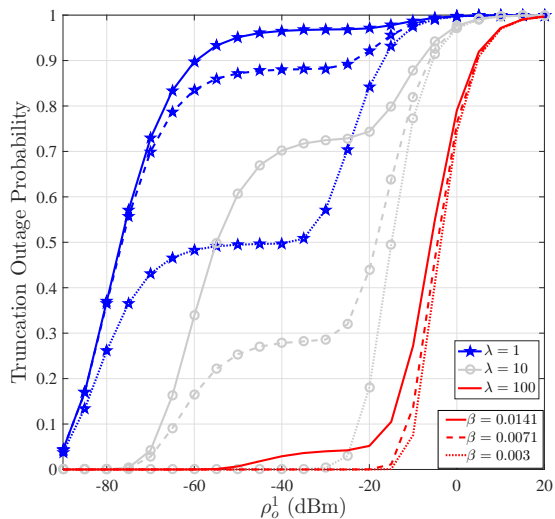} 
 \vspace{-4mm}
  \caption{\small  Effect of the blockage parameter on the truncation outage probability in mmWave network for $\alpha_L^1=2$,  $\alpha_N^1=4$,   BS density $\lambda=1,10$ and $100~\mathrm{BS/km^2}$. \label{fig:Trunc_out_effect_of_beta}}
   \end{center} 
   \vspace{-8mm} 
     \end{figure}

Fig. \ref{fig:Trunc_out_effect_of_beta} shows the effect of blockages on the truncated outage probability. As the average line of sight, which is proportional to the inverse of the blockage parameter value $\frac{1}{\beta}$, increases the truncation outage probability reduces as much lower transmit power is required to meet the receiver sensitivity requirement when the density of blockages and the average size of blockages are much lower. These results are  also in line with the insights previously drawn on equations (\ref{eq:Gamma}) and (\ref{eq:Trun_Outage1}).

\subsubsection{Total Outage Probability}
  \begin{figure}
 \begin{center}  
 \includegraphics[width = .5\textwidth]{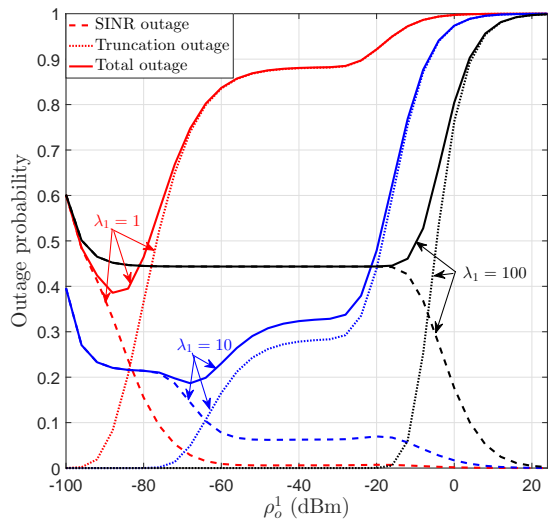} 
 \vspace{-4mm}
  \caption{\small  Total outage probability in a single tier mmWave network with $\theta_1=20$ dB, $\beta_1= 0.0071$ $\alpha_L^1=2$, $\alpha_N^1=4$ and   BS density $\lambda=1,10$ and $100~\mathrm{BS/km^2}$. \label{fig:Total_Outage}}
   \end{center} 
   \vspace{-10mm}
       \end{figure}
In Fig. \ref{fig:Total_Outage}, we show the tradeoff introduced by the cutoff threshold $\rho_o^1$ on the total outage probability, which is defined as $\mathcal O_t=\mathcal O_p +(1-\mathcal O_p)\mathcal O_s$ \cite{ElSawy2014}, for the single tier mmWave network. It can be observed that $\rho_o^1$ tunes the tradeoff between the truncation and SINR outage probabilities and there exists a cutoff threshold $\rho_o^{1\star}$ that minimizes the total outage probability in the single-tier network. Further, the SINR probability dominates the total outage probability at lower cut-off threshold while the truncation outage probability dominates the total outage probability at high values of the cutoff threshold $\rho_o^1$.

\subsubsection{Dense Network Simulation}
 \begin{figure}
 \begin{center}  
 \includegraphics[width = .5\textwidth]{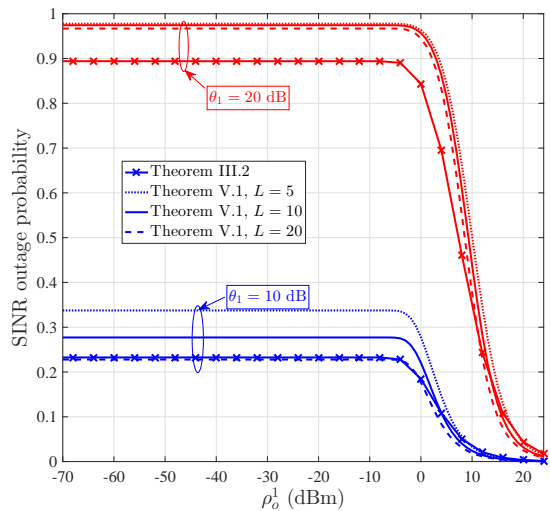} 
    \vspace{-4mm}
  \caption{\small Dense network approximation of the SINR outage probability in a single-tier  mmWave network with    $\alpha_L^1=2$, $\alpha_N^1=4$, $R_B=200~\mathrm m$ and $\lambda_0=100$. \label{fig:Ultra_dense_single_tier}}
   \end{center} 
      \vspace{-10mm}
        \end{figure}
\begin{figure}
 \begin{center}  
 \includegraphics[width = .5\textwidth]{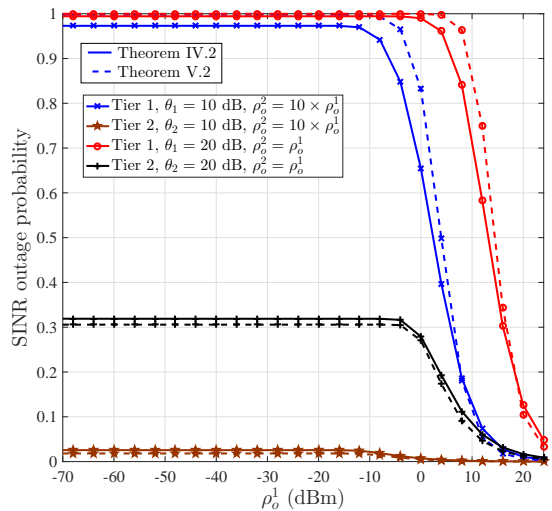} 
  \vspace{-4mm}
  \caption{\small  Dense network approximation of the SINR outage probability in a two-tier  mmWave network with   $\alpha_L^1=2$, $\alpha_N^1=4$, $\alpha_L^2=2.9$, $\alpha_N^2=5$, $R_B=200~\mathrm m$, $\lambda_0=100$, $\lambda_2=2\lambda_1$, $G_{b1}^{\max}=10~\mathrm{dB}$ and  $G_{b2}^{\max}=7~\mathrm{dB}$. \label{fig:Ultra_dense_two_tier}}
   \end{center} 
    \vspace{-10mm}
        \end{figure}

Figs. \ref{fig:Ultra_dense_single_tier} and \ref{fig:Ultra_dense_two_tier} shows the numerical results based on the dense network approximations in Section \ref{Sec:Dense}. In particular, Fig. \ref{fig:Ultra_dense_single_tier} compares the dense network approximation of the SINR outage probability of a single-tier mmWave network given in Theorem \ref{Pr:SINR_ultra} with the exact expression in Theorem \ref{Pr:SINR}. For the dense network approximation, we take the radius of the LOS disc $R_B$ to be equal to $200~\mathrm m$ and a relative BS density $\lambda_0=100$ where $\lambda_0=\lambda_1\pi R_B^2$. The dense network approximation becomes more accurate as $L$ increases. 
Further, Fig. \ref{fig:Ultra_dense_two_tier}, compares the multi-tier dense network approximation in Theorem \ref{Pr:SINR_ultra_Multi_Tier} with the exact expression in Theorem \ref{Theorem:SINR_Multi} while focusing on a two-tier network. The first tier's BS density is obtained from $\lambda_0=\lambda_1\pi R_B^2$ while the second tier's BS density $\lambda_2=2\lambda_1$ and $L$  equals $10$. It can be seen that similar to the single-tier network, the dense network approximation of the SINR outage probability is also fairly accurate for the multi-tier network.


\subsubsection{Average User Transmit Power}
\begin{figure}
 \begin{center}  
 \includegraphics[width = .5\textwidth]{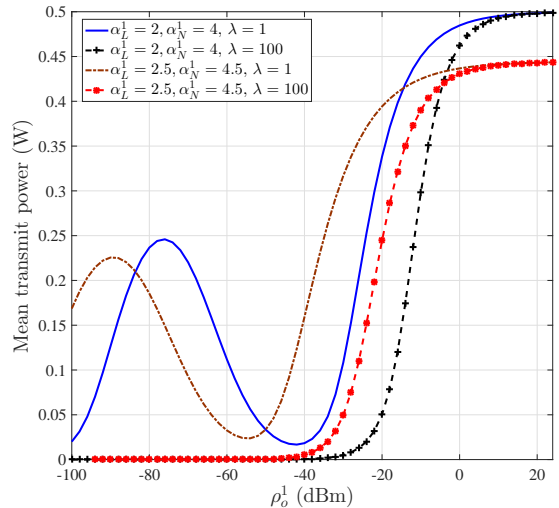} 
  \vspace{-4mm}
  \caption{\small Average transmit power in a single tier mmWave network. \label{fig:Mean_transmit_power}}
   \end{center} 
   \vspace{-10mm} 
     \end{figure}

In Fig. \ref{fig:Mean_transmit_power}, we plot the average transmit power of the users against the cutoff threshold $\rho_0^1$ for the single-tier mmWave network.   It can be observed that for the case with BS density $\lambda=1$, the average transmit power increases with the cutoff threshold for $\rho_o^1$ ranging from $-100$ to $-75$ dBm and it then falls for   $\rho_o^1$ ranging from $-75$ to $-40$ dBm. Afterwards average transmit power then rises with the cutoff threshold until its saturation. Note that increasing $\rho_o^1$ the user need to transmit a higher power to invert the pathloss and maintain a high threshold at the serving BS \cite{ElSawy2014}. However, each user is constrained to a maximum power $P_u$. Hence, a user becomes inactive when its transmit power requirement exceeds $P_u$. An initial increase in $\rho_o^1$ increases the transmit power of all users and hence the first increase in the mean transmit power. A point is reached where the density of active NLOS users starts to decrease  with increasing cutoff threshold since the maximum power constraint cannot  be satisfied and hence the reduction in the mean transmit power. The large discrepancy between the pathloss exponent of the LOS and NLOS users also means a large difference in the transmit power. However, a cutoff threshold is reached where the active LOS user starts to dominate since most NLOS users are inactive, thus leading to an increase in the mean transmit power till its saturation value given bg $\displaystyle\lim_{\rho_o^1\rightarrow\infty}\mathbb E[P_k]=\frac{2}{\min(\alpha_L^1,\alpha_N^1)+2}P_u$.
For dense deployment such as $\lambda=100$,  most of the paths are LOS and the transmit power is non-decreasing with $\rho_o$ in this case.

%
%

 \vspace{-2mm}
\section{Conclusions}\label{Sec:Conclusion}
In this paper, we have presented a stochastic geometry based framework to analyze the SINR outage probability in the uplink of both the single and  multi-tier mmWave cellular networks with truncation channel inversion power control. The framework incorporates the effect of blockages, the per-user  power control as well as the maximum power limitations of the users. Further, each user  controls its transmit power such that the received signal at its serving BS is equal to predefined cutoff threshold. Based on the proposed framework, we derived accurate expressions of the truncation outage probability and  SINR outage probability for the uplink of both the single and  multi-tier mmWave cellular networks. 
 Numerical results show that contrary to the conventional ultra-high-frequency networks there exists a slow growth region for the truncated outage probability. Furthermore, increasing the cutoff threshold does not necessarily lead to a reduction in the SINR outage probability of the mmWave networks.



 \vspace{-2mm}
\appendix

 \vspace{-2mm}
\subsection{Proof of Theorem \ref{lemma:PDF}}\label{Appen:Proof_PDF}
Given that the pathloss exponent $\alpha$ is a random variable which takes the values $\alpha_L$ and $\alpha_N$ with probability $e^{-\beta r}$ and $1-e^{-\beta r}$, respectively, where $r$ is the length of the link. Because the transmit power of a typical user $P_s=\rho_o r_{s}^{\alpha_s}$ for $s=\{L,N\}$, $r_s\in\Phi$ is a function mapping $r_s$ in $\mathbb{R}^2$ to $P_s$ in $\mathbb R$, according to the Mapping theorem \cite[Thm 2.34]{Haenggi2013}, $P_s$ forms an inhomogeneous PPP with the intensity measure $\Lambda(P)=\Lambda_L(P) + \Lambda_N(P)$, where
 \begin{equation}
 \Lambda_L(P)=\frac{2\pi\lambda}{\beta^2}\left(1-e^{-\beta\left(\frac{P}{\rho_o}\right)^{\frac{1}{\alpha_L}}}\left(1+\beta\left(\frac{P}{\rho_o}\right)^{\frac{1}{\alpha_L}}\right)\right),
 \end{equation}
 \begin{equation}
 \Lambda_N(P)=\pi\lambda\left(\frac{P}{\rho_o}\right)^{\frac{2}{\alpha_N}}-\frac{2\pi\lambda}{\beta^2}\left(1-e^{-\beta\left(\frac{P}{\rho_o}\right)^{\frac{1}{\alpha_N}}}\left(1+\beta\left(\frac{P}{\rho_o}\right)^{\frac{1}{\alpha_N}}\right)\right)
 \end{equation}
 \cite{Maamari2016} and the intensity function $\lambda(P)=\frac{\partial \Lambda(P)}{\partial P}=\lambda_L(P)+\lambda_N(P)$, where
  \begin{equation}
  \lambda_L=\frac{2\pi\lambda}{\alpha_L\rho_o^{\frac{2}{\alpha_L}}}P^{\frac{2}{\alpha_L}-1}e^{-\beta\left(\frac{P}{\rho_o}\right)^{\frac{1}{\alpha_L}}}
  \end{equation} and
  
   \begin{equation}
  \lambda_N=\frac{2\pi\lambda}{\alpha_N\rho_o^{\frac{2}{\alpha_N}}}P^{\frac{2}{\alpha_N}-1}\left(1-e^{-\beta\left(\frac{P}{\rho_o}\right)^{\frac{1}{\alpha_N}}}\right).
  \end{equation}
 Therefore, the PDF of $P$ is given by
 \begin{equation}
f_{P}(p)=\frac{\lambda(p)e^{-\Lambda (p)}}{\int_0^{P_u}\lambda(y)e^{-\Lambda (y)}\mathrm dy},~~~~ 0\le p\le P_u
\end{equation}
The PDF of $P$ has been normalized as a result of the truncated channel inversion power control. Furthermore, the $\eta^{th}$ moment of $P$ is thus given by $\int_0^{P_u}p^{\eta}f_P(p)$ and the theorem is obtained.
  \vspace{-2mm}
  \subsection{Proof of Theorem \ref{Pr:SINR} }\label{Appen:Proof_SINR}

Noting that the average interference received from any interfering user (normalized by $\mathcal G$) is strictly less than $\rho_o$. Consequently, the sum interference received at the reference BS from LOS interferers can be expressed from (\ref{eq:SINR}) as
\begin{equation}
I_L =\!\!\!\!\!\! \sum_{u_z\in\Phi_L\backslash\{o\}}\!\!\!\!\!\!\one \!\!\left({P_z\|u_z\|^{-\alpha_L}}<\rho_o\!\right)\!\!P_zG_z|g_z|^2|u_z\|^{-\alpha_L}
\end{equation}
where $\Phi_L$ is a PPP of LOS interferers,  and $\one(.)$ is an indicator function which takes the values of one when $(.)$ is true and zero otherwise. Consequently, the Laplace transform of the aggregate interference from LOS interferer $\mathcal L_{I_L}$ in (\ref{eq:Prob1}) can be computed as
\begin{align}
\mathcal L_{I_L}&= \mathbb E_{\Phi_L}\left[e^{-snI_L}\right]\\
&=\!\mathbb E_{\Phi_L}\!\!\left[e^{-sn\sum_{u_z\in\Phi_L\backslash\{o\}}\!\!\one\! \left({P_z\|u_z\|^{-\alpha_L}}<\rho_o\!\right)\!P_zG_z|g_z|^2|u_z\|^{-\alpha_L}}\right]\nonumber\\
&\overset{(b1)}=\!\mathbb E_{P_z,g_z,G_z}\!\!\!\left[\!\prod_{u_z\in\Phi_L\backslash\{o\}}\!\!\!\!\!\!\!\!\!e^{-sn\one\left(\!\|u_z\|>\left(\!\frac{P_z}{\rho_o}\!\right)^\frac{1}{\alpha_L}\!\right)\!P_zG_z|g_z|^2|u_z\|^{-\alpha_L}}\right]\nonumber\\
&\overset{(b2)}=e^{\left(\!-2\pi\lambda\sum_{v=1}^4\!b_v\!\int_{\left(\frac{P}{\rho_o}\right)^\frac{1}{\alpha_L}}^\infty \!\!\mathbb E_{P,g}\left[\left(1-e^{-sna_vPgr^{-\alpha_L}}\right)\right]re^{-\beta r}\mathrm dr\right)}
\nonumber\\
&\overset{(b3)}=e^{\!\left(\!-2\pi\lambda\!\sum_{v=1}^4\!b_v\!\!\int_{\left(\frac{P}{\rho_o}\right)^\frac{1}{\alpha_L}}^\infty \!\!\!\mathbb E_{P}\!\left[\!\left(\!1-\!\frac{1}{\left(1+sna_vPr^{-\alpha_L}/N\right)^N}\!\right)\!\right]re^{-\beta r}\mathrm dr\!\right)}\nonumber\\
&=\prod_{v=1}^4e^{-2\pi\lambda b_v\int_{\left(\frac{P}{\rho_o}\right)^\frac{1}{\alpha_L}}^\infty \mathbb E_{P}\left[\left(1-\frac{1}{\left(1+sna_vPr^{-\alpha_L}/N\right)^N}\right)re^{-\beta r}\right]\mathrm dr}\nonumber
\end{align}
\begin{align}
&\overset{(b4)}=\!\!\prod_{v=1}^4\!\!e^{\!-2\pi\lambda q_v^{\frac{2}{\alpha_L}} \!b_v\!\int_{\mathcal{A}}^\infty\!\!\!\int_{0}^{P_u}\!\left(\!\!1\!-\frac{1}{\left(\!1+\frac{y^{-\alpha_L}}{N}\right)^N}\!\!\right)\!yP^{\frac{2}{\alpha_L}}\!e^{-\beta \left(q_vP\right)^{\frac{1}{\alpha_L}}\!\!y}\!f_P\mathrm dP\mathrm dy}\nonumber\\
&=e^{-Q_n}\nonumber,
\end{align}
where $(b1)$ follows from the independence of $\Phi_L$, $g_z$, $G_z$ and $P_z$, $(b2)$ follows from the probability generation functional (PGFL) of the PPP \cite{Andrews2011} and  the independence of the interference link directivity gain $G_z$
with probability distribution $G_z = a_v$ with probability $b_v$, $(b3)$ follows from from computing the moment generating function of a gamma random variable $g$, $(b4)$ is obtained by changing the variables $y = {r}/{\left(sna_vP\right)^{\frac{1}{\alpha_L}}}$ while $f_P$ is given in (\ref{eq:PDF}). Further,  $\mathcal{A}=\left(sna_v\rho_o\right)^{-\frac{1}{\alpha_L}}$  and $q_v=sna_v$. 
Similarly, the for the  NLOS interfering links, $\mathcal L_{I_N}$can be computed as
\begin{align}
\mathcal L_{I_N}&=\mathbb E_{\Phi_N}\left[e^{-snI_N}\right] \\
&=\!\prod_{v=1}^4\!e^{\!-2\pi\lambda q_v^{\frac{2}{\alpha_N}}\! b_v\!\int_{\mathcal B}^\infty\!\!\int_{0}^{P_u}\!\left(\!1-\frac{1}{\left(1+\frac{y^{-\alpha_N}}{N}\right)^N}\!\right)\!yP^{\frac{2}{\alpha_N}}Z(y,p)f_P\mathrm dP\mathrm dy}\nonumber\\
&=e^{-V_n}\nonumber,
\end{align}
where $\mathcal B=\left(sna_v\rho\right)^{-\frac{1}{\alpha_N}}$ and $Z(y,p)=\left(1-e^{-\beta \left(q_vP\right)^{\frac{1}{\alpha_N}}y}\right)$. 
 \vspace{-2mm}
\subsection{Proof of Theorem \ref{Lemma:PDF_Multi}}
\label{Appen:Proof_PDF_Multi}
Given that $y_k=\displaystyle\min_{m_k\in\Phi _k}(||u-m_k||^{\alpha_s^k}) $ is used to select the serving BS in the $k^{th}$ tier. 
 Then $f_{y_k}(y)=\lambda_k(y)e^{-\Lambda_k(y)}$ where 
\begin{align}
\lambda_k(y) = \frac{2\pi\lambda_k}{\alpha_L^k }y^{\frac{2}{\alpha_L^k}-1}e^{-\beta_k{y}^{\frac{1}{\alpha_L^k}}}+\frac{2\pi\lambda_k}{\alpha_N^k }y^{\frac{2}{\alpha_N^k}-1}\left(1-e^{-\beta_k{y}^{\frac{1}{\alpha_N^k}}}\right)
\end{align}
and
\begin{align}\label{eq:Lambda_multi_tier2}
\Lambda_k(y)=\frac{2\pi\lambda_k}{\beta_k^2}\left(1-e^{-\beta_k{y}^{\frac{1}{\alpha_L^k}}}\left(1+\beta_k{y}^{\frac{1}{\alpha_L^k}}\right)\right)+\pi\lambda_k{y}^{\frac{2}{\alpha_N^k}}
-\frac{2\pi\lambda_k}{\beta_k^2}\left(1-e^{-\beta_k{y}^{\frac{1}{\alpha_N^k}}}\left(1+\beta_k{y}^{\frac{1}{\alpha_N^k}}\right)\right)
.\end{align}
Noting that the user connects to the BS that provides the
maximum average received signal, we can follow the same approach for the multi-tier UHF network in \cite{ElSawy2014}. The typical user $u$ connects to the reference BS  from the $j^{th}$ tier, then $y_j=\displaystyle\min_{k}(y_k)$. The transmit power of the typical user connected to the reference user in the $j^{th}$ tier is given by $P_j=\rho_o^j\displaystyle\min_{k}(y_k)$ where $P_j\le P_u$. Consequently, the cumulative distribution function (CDF) of the transmit power can be expressed as
\begin{equation}
F_{P_j}(p) =\frac{1-e^{-\sum_{k=1}^K{ \Lambda_k\left(\frac{p}{\rho_o^j}\right)}}}{1-e^{-\sum_{k=1}^K{ \Lambda_k\left(\frac{P_u}{\rho_o^j}\right)}}}
\end{equation}
and the PDF of the transmit power is given as
\begin{align}
f_{P_j}(p)&=\frac{\mathrm dF_{P_j}(p)}{\mathrm{d}p}\\
&=\frac{\sum_{k=1}^{K}\overline{\lambda}_k(p)}{1-e^{-\sum_{a=1}^K{ \Lambda_a\left(\frac{P_u}{\rho_o^j}\right)}}}{e^{-\sum_{b=1}^K{ \Lambda_b\left(\frac{p}{\rho_o^j}\right)}}}\nonumber
\end{align}
where
\begin{equation}
\overline{\lambda}_k(p) =\frac{2\pi\lambda_k}{\alpha_L^k {\rho_o^j}^{2/\alpha_L^k}}y^{\frac{2}{\alpha_L^k}-1}e^{-\beta_k\left(\frac{y}{\rho_o^j}\right)^{\frac{1}{\alpha_L^k}}}+\frac{2\pi\lambda_k}{\alpha_N^k {\rho_o^j}^{2/\alpha_N^k}}y^{\frac{2}{\alpha_N^k}-1}\left(1-e^{-\beta_k\left(\frac{y}{\rho_o^j}\right)^{\frac{1}{\alpha_N^k}}}\right)
\end{equation}
and $\Lambda_c(.)$ is given in (\ref{eq:Lambda_multi_tier2})
 \vspace{-2mm}
\subsection{Proof of Theorem \ref{Theorem:SINR_Multi}}
\label{Appen:Proof of SINR_Multi}
Noting that the average interference received from any interfering user from the $k^{th}$ tier  is less than $\rho_o^k$. The sum interference received at the reference BS in  the $j^{th}$ tier from  LOS users in the $k^{th}$ can be expressed as
\begin{equation}
I_L^k=\sum_{u_z\in\Phi_L^k\backslash\{o\}}{\one\left(P_{zk}||u_z||^{-\alpha_L^j}<\rho_o^k\right)P_{zk}G_{z}|g_z|^2||u_z||^{-\alpha_L^j}},
\end{equation}
where $\Phi_L^k$ is a PPP of LOS interfering users from the $k^{th}$ tier. The indicator function is used to capture the correlation among the location of the interfering users and the location of the reference BS. Hence, the Laplace transform of the aggregate interference from LOS users in the $k^{th}$ tier received by the reference mmWave BS in the $j^{th}$ tier $\mathcal L_{I_L^k}$ can be computed as
\begin{align}
\mathcal L_{I_L^k}&=\mathbb E_{\Phi_L^k}\left[e^{-snI_L^k}\right]\\&=\mathbb E_{\Phi_L^k}\left[e^{-sn\sum_{u_z\in\Phi_L^k\backslash\{o\}}{\one\left(P_{zk}||u_z||^{-\alpha_L^j}<\rho_o^k\right)P_{zk}G_{z}|g_z|^2||u_z||^{-\alpha_L^j}}}\right]\nonumber
\end{align}
\begin{align}
&=\exp\left(-2\pi\lambda_k\sum_{v=1}^4b_v^j\int\limits_{\left(\frac{P_k}{\rho_o^k}\right)^{\frac{1}{\alpha_L^j}}}^{\infty}\mathbb E_{P_k,g}\left[\left(1-e^{-sna_v^jP_kgr^{-\alpha_L^j}}\right)\right]re^{-\beta_jr}\mathrm dr \right)\nonumber\\
&=\exp\left(-2\pi\lambda_k\sum_{v=1}^4b_v^j\int\limits_{\left(\frac{P_k}{\rho_o^k}\right)^{\frac{1}{\alpha_L^j}}}^{\infty}\mathbb E_{P_k}\left[\left(1-\frac{1}{\left(1+sna_v^jP_kr^{-\alpha_L^j}/N\right)^N}\right)re^{-\beta_jr}\right]\mathrm dr \right)\nonumber\\
&=e^{\left(-2\pi\lambda_k\sum_{v=1}^4b_v^j{q_v^j}^{\frac{2}{\alpha_L^j}}\int\limits_{\left(sna_v^j\rho_o^k\right)^{-\frac{1}{\alpha_L^j}}}^{\infty}\int\limits_0^{P_u}\left(1-\frac{1}{\left(1+\frac{y^{-\alpha_L^j}}{N}\right)^N}\right)y P_k^{\frac{2}{\alpha_L^j}}e^{-\beta_j\left(q_v^jP_k\right)^{\frac{1}{\alpha_L^j}}y}f_{P_k}\mathrm dP\mathrm dy\right)}\nonumber\\
&=e^{-Q_n^k}\nonumber
\end{align}
Similarly, the Laplace transform of the aggregate interference from NLOS users in the $k^{th}$ tier received by the reference BS in the $j^{th}$ tier $\mathcal L_{I_N^k}$ can be expressed as
\vspace{-2mm}
\begin{align}
\mathcal L_{I_N^k}&=\mathbb E_{\Phi_N^k}\left[e^{-snI_N^k}\right]\\
&=e^{\left(-2\pi\lambda_k\sum_{v=1}^4b_v^j{q_v^j}^{\frac{2}{\alpha_N^j}}\int\limits_{\left(sna_v^j\rho_o^k\right)^{-\frac{1}{\alpha_N^j}}}^{\infty}\int\limits_0^{P_u}\left(1-\frac{1}{\left(1+\frac{y^{-\alpha_N^j}}{N}\right)^N}\right)y P_k^{\frac{2}{\alpha_N^j}}\left(1-e^{-\beta_j\left(q_v^jP_k\right)^{\frac{1}{\alpha_N^j}}y}\right)f_{P_k}\mathrm dP\mathrm dy\right)}=e^{-V_n^k}\nonumber
\end{align}
 \vspace{-12mm}
\subsection{Proof of Theorem \ref{Pr:SINR_ultra}}\label{Appen:Proof_SINR_Ultra}
\vspace{-1mm}
The proof is based on the key assumptions in Appendix \ref  {Appen:Proof_SINR} such that
\vspace{-1mm}
\begin{align}\label{eq:outage_dense2}
&\mathcal{L}_{I_L^k}=\mathbb{E}_{\Phi_L^k}\left[-slI_L^k\right]
=\mathbb E_{\Phi_L^k}\left[e^{-sl\sum_{u_z\in\Phi_L^k\cap\mathcal B(0, R_B)\backslash\{o\}}{\one\left(P_{zk}||u_z||^{-\alpha_L^j}<\rho_o^k\right)P_{zk}G_{z}||u_z||^{-\alpha_L^j}}}\right]\nonumber\\
&\overset{(e1)}=\exp\left(\mathbb{E}_{P_k}\left[\sum_{v=1}^{4}-2\pi\lambda_k b_v^j\int_{\left(\frac{P_k}{\rho_o^k}\right)^{\frac{1}{\alpha_L^j}}}^{R_B}\left(1-e^{slP_ka_v^jr^{-\alpha_L^j}}\right)r\mathrm dr\right]\right)\nonumber\\
&\overset{(e2)}=\exp\left(\mathbb{E}_{P_k}\left[\pi\lambda_k\left(\left(\frac{P_k}{\rho_o^k}\right)^{\frac{2}{\alpha_L^j}}-R_B^2\right)+\sum_{v=1}^{4}\frac{2\pi\lambda_k b_v^j}{\alpha_L^j}\left(sla_L^j\right)^{\frac{2}{\alpha_L^j}}P_k^{\frac{2}{\alpha_L^j}}\int_{sla_v^jP_kR_B^{-\alpha_L^j}}^{sla_v^j\rho_o^k}\frac{e^{-w}}{w^{1+\frac{2}{\alpha_L^j}}}\mathrm dv\right]\right)\nonumber\\
&\overset{(e3)}=e^{\left(\int_{0}^{P_u}\left(\pi\lambda_k\left(\left(\frac{P_k}{\rho_o^k}\right)^{\frac{2}{\alpha_L^j}}-R_B^2\right)+\sum_{v=1}^{4}\frac{2\pi\lambda_k b_v^j}{\alpha_L^j}
\left(sla_L^j\right)^{\frac{2}{\alpha_L^j}}P_k^{\frac{2}{\alpha_L^j}}\left(\Gamma\left(\frac{-2}{\alpha_L^j},\frac{\eta l\theta_j a_v^jP_k}{\rho_o^j\mathcal G_jR_B^{\alpha_L^j}}\right)-\Gamma\left(\frac{-2}{\alpha_L^j},\frac{\eta l\theta_j a_v^j}{\mathcal G_j}\right)\right)
\right)f_{P_k}\mathrm dP_k\right)},
\end{align}
where $(e1)$ follows from computing the PGFL of the PPP $\Phi_L^k$, $(e2)$ follow from a change of variable $w=sla_v^jP_kr^{-\alpha_L^j}$ and $(e3)$ follows from the simplified  PDF of the transmit power over the LOS region. The PDF of the uplink transmit power in the $k^{th}$ tier of dense mmWave networks can be obtained by noting that $P_k=\rho_o^k r_k^{\alpha_L}$, where $r_k$
is the $k^{th}$ tier uplink distance in the dense deployment which follows a Rayleigh distribution $f_{r_k}(r)=2\pi\lambda r e^{-\pi\lambda r^2},~0\le r\le\infty$. Consequently, following the same approach in Appendix \ref{Appen:Proof_PDF_Multi} we obtain the PDF of the $k^{th}$ tier transmit power as
\vspace{-2mm}
  \begin{equation}
  f_{P_j}(x)=\frac{\sum_{k=1}^K\frac{2\pi\lambda_k p^{\frac{2}{\alpha_L^k}-1}}{\alpha_L^k\left(\rho_o^j\right)^{\frac{2}{\alpha_L^k}}}}{1-e^{-\sum_{a=1}^K\pi\lambda_a\left(\frac{P_u}{\rho_o^j}\right)^{\frac{2}{\alpha_L^a}}}}e^{-\sum_{b=1}^K\pi\lambda_b\left(\frac{p}{\rho_o^j}\right)^{\frac{2}{\alpha_L^b}}}
 \end{equation}
 
 Finally, the approximation of the SINR outage probability in the uplink of a multi-tier mmWave network given in (\ref{eq:SINRoutage_dense}) can be obtained by substituting (\ref{eq:outage_dense2}) for (\ref{eq:outage_dense1}) and with $\lambda_0 =\lambda\pi R_B^2$.
\vspace{-5mm}
 \ifCLASSOPTIONcaptionsoff
  \newpage
\fi
\bibliographystyle{IEEEtran}
\bibliography{IEEEabrv,EE_SE_Trunc}

\end{document}